\documentclass[11pt]{article}
\usepackage{amsmath,amssymb,amsfonts,color,epsf}
\usepackage{amssymb,slashed,latexsym,multirow,color}
\usepackage{graphics}
\makeatletter
\@addtoreset{equation}{section}
\makeatother

\newcommand{\hoch}[1]{$\, ^{#1}$}

\newcommand{\be}{\begin{equation}}
\newcommand{\ee}{\end{equation}}
\newcommand{\bea}{\setlength\arraycolsep{2pt} \begin{eqnarray}}
\newcommand{\eea}{\end{eqnarray}}

\usepackage[font=footnotesize,labelsep=newline,labelfont=sc,justification=centering,position=top]{caption}

\usepackage{hyperref}

\newsavebox{\uuunit}
\sbox{\uuunit}
    {\setlength{\unitlength}{0.825em}
     \begin{picture}(0.6,0.7)
        \thinlines
        \put(0,0){\line(1,0){0.5}}
        \put(0.15,0){\line(0,1){0.7}}
        \put(0.35,0){\line(0,1){0.8}}
       \multiput(0.3,0.8)(-0.04,-0.02){12}{\rule{0.5pt}{0.5pt}}
     \end {picture}}

\def\be{\begin{equation}}
\def\ee{\end{equation}}
\def\ba{\begin{array}}
\def\ea{\end{array}}
\def\bea{\begin{eqnarray}}
\def\eea{\end{eqnarray}}
\def\bd{\begin{displaymath}}
\def\ed{\end{displaymath}}

\def\t{\tau}

\begin{document}

\begin{flushright}
\hfill{ \
\ \ \ \ UG-2016-19 \ \ \ \ \\
\ \ \ \ ICCUB-16-026 \ \ \ }
\end{flushright}
\vskip 1.2cm

\begin{center}
{\Large \bf
{
Carroll versus Galilei Gravity}}
\end{center}
\vspace{25pt}
\begin{center}
{\Large {\bf }}

\vspace{10pt}

{\Large Eric Bergshoeff\hoch{*
}\,,
 Joaquim Gomis\hoch{2}, Blaise Rollier\hoch{*
}\,, Jan Rosseel\hoch{3} and Tonnis ter Veldhuis\hoch{*}\footnote{On leave of absence from Macalester College, Saint Paul (USA).}
 }

\vspace{10pt}

\hoch{*} {\it Centre for Theoretical Physics, University of Groningen,\\
Nijenborgh 4, 9747 AG Groningen, The Netherlands}\\

 \hoch{2} {\it Departament de F\'{\i}sica Cu\`antica i Astrof\'{\i}sica and Institut de Ci\`encies del\\
Cosmos, Universitat de Barcelona, Mart\'i i Franqu\`es 1, E-08028 Barcelona, Spain}

\hoch{3} {\it Faculty of Physics, University of Vienna,\\
Boltzmanngasse 5, A-1090, Vienna, Austria
}

\vspace{30pt}

\underline{ABSTRACT}
\end{center}

We consider two distinct limits of General Relativity that in contrast to the standard non-relativistic limit can be taken at the level of the Einstein-Hilbert action
instead of the equations of motion. One is a non-relativistic limit and leads to a so-called Galilei gravity theory, the other is an ultra-relativistic limit yielding a so-called Carroll gravity theory. We present both gravity theories in a first-order formalism and show that in both cases the equations of motion (i) lead to constraints on the geometry
and (ii) are not sufficient to solve for all of the components of the connection fields in terms of the other fields.  Using a second-order formalism we show that these
independent components serve as Lagrange multipliers for the geometric constraints we found earlier. We point out a few noteworthy differences between Carroll
and Galilei gravity and give some examples of matter couplings.

\vspace{15pt}

\thispagestyle{empty}

\vspace{15pt}

 \vfill

\thispagestyle{empty}
\voffset=-40pt

\newpage

\tableofcontents


\newpage


\section{Introduction}

Einstein's classical theory of General Relativity  is able to explain many experiments within certain distance scales. However, it is generally appreciated that there are issues both at small distances where the unification of General Relativity with quantum mechanics becomes relevant as well as at large distances where gravity may couple to as yet un-seen dark matter and where we are facing the dark energy puzzle.
A remarkable result of the quest for a theory of quantum gravity is the AdS/CFT correspondence \cite{Maldacena:1997re,Gubser:1998bc,Witten:1998qj}
which states that a gravitational theory in a $D$-dimensional
Anti-de Sitter (AdS) spacetime under certain conditions can be described by a relativistic Conformal Field Theory (CFT) that is defined at the boundary of that spacetime.

The AdS/CFT correspondence has been generalized to a non-relativistic correspondence where one considers gravitational background solutions in the bulk that preserve a number of non-relativistic symmetries such as the Schr\"odinger symmetries \cite{Son:2008ye, Balasubramanian:2008dm}
 or Lifshitz symmetries \cite{Kachru:2008yh}. There exists another approach, initiated in \cite{Bagchi:2009my},
where not only the boundary QFT is non-relativistic but also the String Theory. Non-relativistic strings came into the picture some time ago as a possibly solvable special sector of String Theory \cite{Gomis:2000bd,Danielsson:2000gi}. In this alternative approach  one ends up with a
non-relativistic vibrating string in the bulk  \cite{Gomis:2005pg}. When the curvature is small the non-relativistic string gives rise to  a non-relativistic gravity
theory in the bulk with a two-dimensional foliation, representing the time and the single spatial direction of the string.
This gravity theory is a string-like version of a  frame-independent formulation of Newton's theory of gravity, called Newton-Cartan (NC) gravity,  which has a one dimensional foliation representing the absolute time.

In view of its role in the AdS/CFT correspondence,  it is of interest to consider
 special limits of General Relativity, possibly with matter beyond the standard non-relativistic limit which gives rise to NC gravity.\,\footnote{We will not consider in this paper the Newtonian limit, which is discussed in most text books, since that limit involves extra assumptions leading to a frame-dependent formulation.}
 Motivated by this we will consider in this paper two distinct limits of General Relativity with a one-dimensional foliation. The extension to a two-dimensional foliation can be done in a separate step and will not be considered in this paper.


The standard non-relativistic limit  of General Relativity in four spacetime dimensions, leading to NC gravity, that is usually considered in the literature can only be defined at the level of the equations of motion.\,\footnote{In three dimensions the non-relativistic limit has been considered at the level of the action by adding an extra term to the Einstein-Hilbert action \cite{Bergshoeff:2016lwr}.}
This
 so-called NC limit leads to infinities when applied at the level of the Einstein-Hilbert (EH) action. A noteworthy feature of the resulting NC gravity theory is that it contains a central charge gauge field that couples to the current corresponding to the conservation of (massive) particles.

In this paper we will explore two different limits of General Relativity that, in contrast to the NC  limit, can be defined at the level of the EH action.  The first limit we will consider is an ultra-relativistic limit leading to a so-called Carroll gravity theory invariant under reparametrizations and the Carroll symmetries.\,\footnote{
 A different version of Carroll gravity has been studied in \cite{Hartong:2015xda}. We will compare the two versions later in this paper.}
These Carroll symmetries have recently occurred in studies of flat space holography \cite{Banks:2016xsj}.
The second limit that we will consider  is a non-relativistic limit, the so-called Galilei limit, that differs from the NC limit in the sense that it does not involve a mass parameter and a central charge gauge field. The resulting Galilei gravity theory is invariant under reparametrizations and Galilei symmetries. Such symmetries, and extensions thereof, have occurred in a recent study on non-relativistic limits of string actions \cite{Batlle:2016iel,Gomis:2016zur}.


In this paper we will present the limits of General Relativity leading to the Carroll and Galilei gravity theories using a first-order formulation where the spin-connection fields are considered to be independent variables. A noteworthy feature is that the equations of motion lead to constraints on the geometry.
We next show that, in contrast to General Relativity, for both Carroll and Galilei gravity not all components of the spin-connection fields  can be solved for  by using the equations of motion. Instead, we find that, using a second-order formulation, the independent components of the spin-connection fields, occur as Lagrange multipliers that precisely reproduce  the geometric constraints mentioned above.

The organization of this paper is as follows. In section 2 we review a few aspects of General Relativity
that are relevant for the analysis  in the next sections. In section 3 we explore Carroll gravity, both using a first-order as well as a second-order formulation. In section 4 we perform a similar analysis for Galilei gravity. In section 5 we discuss matter couplings for both Carroll and Galilei gravity. Finally, we give our conclusions in section 6.

\section{General Relativity}

Before taking limits we first summarize some relevant formulae of General Relativity including matter couplings which will be of use in the next sections.
Our starting point is  the $D$-dimensional Poincar\'e algebra of spacetime translations $P_A$ and Lorentz transformations $J_{AB} \ (A=0,1,\dots,D-1)$
\begin{eqnarray}
\left[P_A,J_{BC}\right] &=& 2\eta_{A[C}P_{B]}\,,\label{Poincare1}\\
\left[J_{AB}, J_{CD}\right] &=& 4\eta_{[A[D } J_{C]B]}\,,\label{Poincare2}
\end{eqnarray}
where $\eta_{AB}$ is the (mostly plus) Minkowski metric. To each generator of the Poincar\'e algebra we associate a gauge field, a local parameter parametrizing
the corresponding symmetry and a  curvature, see Table \ref{Poincaretable}. The gauge field $E_\mu{}^A$ is  the Vielbein field while $\Omega_\mu{}^{AB}$
is the spin-connection field.

{\small
\begin{table}[t]
\begin{center}
\begin{tabular}{|c|c|c|c|c|}
\hline
symmetry&generators& gauge field&parameters&curvatures\\[.1truecm]
\hline\rule[-1mm]{0mm}{6mm}
spacetime translations&$P_A$&$E_\mu{}^A$&$\eta^A$&$R_{\mu\nu}{}^A(P)$\\[.1truecm]
Lorentz transformations&$J_{AB}$&$\Omega_\mu{}^{AB}$&$\Lambda^{AB}$&$R_{\mu\nu}{}^{AB}(J)$\\[.1truecm]
\hline
\end{tabular}
\end{center}
\caption{This table indicates the generators of the Poincar\'e algebra and the  gauge fields, local parameters and curvatures that are associated to each of these generators.}\label{Poincaretable}
\end{table}
 }

According to the Poincar\'e algebra \eqref{Poincare1}, \eqref{Poincare2}  the gauge fields transform as follows:\,\footnote{All parameters depend on the coordinates $x^\mu$, even when not explicitly indicated.}
\begin{eqnarray}\label{Ptransf}
\delta E_\mu{}^A  &=& \partial_\mu \eta^A  + \Lambda^A{}_B E_\mu{}^B -   \Omega^{AB}_\mu \eta_B\,, \\[.1truecm]
\delta \Omega_\mu^{AB} &=& \partial_\mu \Lambda^{AB} + \Omega_\mu^A{}_C \Lambda^{BC} - \Omega^{BC}_\mu \Lambda^A{}_C\,.
\end{eqnarray}
These gauge fields transform as covariant vectors under general coordinate transformations with parameters $\xi^\mu$. The curvatures indicated in Table   \ref{Poincaretable} transform covariantly under these transformations:
\begin{eqnarray}
R_{\mu\nu}{}^A(P) &=& 2\partial_{[\mu} E^A_{\nu]} -  2 \Omega^{A}_{[\mu}{}_B E^B_{\nu]}   \,,\label{eq: constraint spin}\\
R_{\mu\nu}{}^{AB}(J) &=& 2\partial_{[\mu} \Omega^{AB}_{\nu]} -  2 \Omega^{BC}_{[\mu} \Omega^{A}_{\nu]}{}_C  \,. \label{eq: constraint MA}
\end{eqnarray}

In arbitrary dimensions, it is not possible to write down a gauge-invariant action for the gauge fields \cite{Banados:1996hi}.  Instead, we consider the following action
which is invariant under general coordinate transformations and local Lorentz transformations:
\begin{equation}
S= - \frac{1}{16\pi G_N} \int E E^\mu_A E^\nu_B R_{\mu\nu}{}^{AB}(J) + S_{\rm matter}\,. \label{eq: ActionS}
\end{equation}
Here $E$=det$E_\mu{}^A$ and we have defined the inverse Vierbein $E^\mu{}_A$
\begin{equation}
E_\mu{}^A E^\mu{}_B = \delta^A_B\,,\hskip 1truecm E_\mu{}^A E^\nu{}_A = \delta_\mu^\nu\,.
\end{equation}
For generality we have included an arbitrary matter action $S_{\rm matter}$. Note that we are using  a first-order formulation where $\Omega_\mu{}^{AB}$ is treated as an independent variable.
The action \eqref{eq: ActionS}  transforms under $P$-transformations as follows:
\begin{equation}
\delta_P S =- \frac{3}{8\pi G_N} \int E E_{[A}^\mu E_B^\nu E_C^\rho E_{D]}^\sigma R_{\mu\nu}{}^{AB}(J) R_{\rho\sigma}{}^C(P) \eta^D + \delta_P S_{\tiny{\mbox{mat}}}\,.
\label{eq: varActionS}
\end{equation}
This shows that only for $D=3$ the gravity kinetic term in the action \eqref{eq: ActionS} is invariant under both Lorentz and $P$-transformations.
This is related to the fact that for $D=3$ this kinetic term  can be rewritten as a Chern-Simons gauge theory.

Varying the action \eqref{eq: ActionS} with respect to the independent gauge fields $\Omega_\mu{}^{AB}$ and $E^\mu{}_A$ we obtain the following equations of motion:
\begin{eqnarray}
R_{C[A}{}^{C}(P)E_{B]}^{\mu} + \frac{1}{2}E_{C}^{\mu}R_{AB}{}^{C}(P) &=&  J^\mu_{AB} \,, \label{eq: varS Omega}\\
G_\mu{}^A &=&  T_\mu{}^A \,, \label{eq: varS E}
\end{eqnarray}
where the Einstein tensor is defined by
\begin{equation}
G_\mu{}^A = R_{\mu C}{}^{A C}(J) - \frac{1}{2}E^{A}_{\mu} R_{CD}{}^{CD}(J)
\end{equation}
and where we have defined the Lorentz transformation current $J^\mu_{AB}$ and the energy-momentum tensor $T_\mu{}^A$ as follows ($\kappa = 8\pi G_N$):
\begin{eqnarray}
J^\mu_{AB} &\equiv & \frac{\kappa}{E}\frac{\delta S_{\tiny{\mbox{mat}}}}{\delta \Omega^{AB}_\mu}  \,,\hskip 1.5truecm
T_\mu{}^A = \frac{\kappa}{E}\frac{\delta S_{\tiny{\mbox{mat}}}}{\delta E^\mu_A}  \,.
\end{eqnarray}
For $D>2$ the equation of motion (\ref{eq: varS Omega}) can be rewritten as
\begin{equation}
R_{\mu\nu}{}^{A}(P) = 2J^\rho_{AB} E^A_\mu E^B_\nu E^{C}_{\rho} + \frac{4}{D-2}J^\rho_{AB} E^A_\rho E^B_{[\mu} E^C_{\nu]} \,.\label{eq: PoincRP zero}
\end{equation}
By taking cyclic permutations, this equation can be further rewritten  in terms of the Lorentz spin connection as follows:
\begin{equation}
\Omega_\mu^{AB} = - 2 E^{\rho [A}\partial_{[\mu} E_{\rho]}^{B]} + E_{\mu C} E^{\rho A} E^{\nu B}\partial_{[\rho} E_{\nu]}^C + X^{AB}_\mu\,, \label{eq: sol relspin connection}
\end{equation}
with
\begin{equation}
X^{AB}_\mu =  2E^{C}_{\mu}\eta^{D[A} J_{C D}^{B]}   - E_{\mu C} \eta^{A E} \eta^{B F} J_{E F}^C   + \frac{4}{D-2} E^{[A}_{\mu} \eta^{B] D}J_{C D}^C \,.
\end{equation}

The equations of motion \eqref{eq: varS Omega} and \eqref{eq: varS E} give relations between the curvatures and the currents.  The curvatures satisfy the following Bianchi identities:
 \begin{eqnarray}
\mathcal{D}_{[\mu} R_{\nu\rho]}{}^A(P) + R_{[\mu \nu}{}^{AB}(J)E_{\rho] B} &=& 0 \,,\label{eq: Bianchi1}\\
\mathcal{D}_{[\mu } R_{\nu \rho]}{}^{AB}(J)  &=& 0 \,, \label{eq: Bianchi2}
\end{eqnarray}
where  $\mathcal{D}_\mu$ is the Lorentz-covariant derivative. By contraction these Bianchi identities imply that
\begin{eqnarray}
2R_{[\mu\nu]} + \mathcal{D}_A R_{\mu\nu}{}^A(P) + 2 \mathcal{D}_{[\mu} R_{\nu]A}{}^A(P)  & = & 0  \,,\label{eq: Bianchicont1} \\
2\mathcal{D}_A G_C{}^A  - 2 R_B{}^{A}R_{CA}{}^B(P) +  R_{CD}{}^{A B}(J) R_{AB}{}^{D}(P) &=& 0 \,,\label{eq: Bianchicont2}
\end{eqnarray}
with
\begin{eqnarray}
R_{\mu}{}^A =   R_{\mu\nu}{}^{AB}(J)E_B^\nu\,.
\end{eqnarray}
For the equations of motion to be consistent, these identities require the following on-shell relations among the currents:
\begin{eqnarray}
T_{[AB]}  &=& - \mathcal{D}_C J^C_{AB}  + \frac{2}{D-2} J^C_{AB} J^D_{CD} \,,\\
\mathcal{D}_B T_A{}^B &=&  2J^\mu_{AB}T_{\mu}{}^{B} - R_{AB}{}^{CD}(J)J_{CD}^B  + \frac{2}{D-2}\left(2T_A{}^{B}  - T \delta_A^B\right)J^C_{BC}  \,.
\end{eqnarray}

\section{Carroll Gravity}\label{sec: Carroll}

In this section we will consider Carroll gravity, i.e.~the ultra-relativistic limit of General Relativity. The underlying algebra is a particular (ultra-relativistic) contraction of the
Poincar\'e algebra which is called the Carroll algebra \cite{Levy-Leblond}.  This section consists of two subsections. In the first subsection we will review a few properties of the Carroll algebra while in the second one we will construct Carroll gravity. The addition of general matter couplings to Carroll gravity will be discussed in subsection \ref{sec: Car matter}.

\subsection{The Carroll Algebra}

The Carroll algebra is obtained by a contraction of the Poincar\'e algebra. To define this contraction, we decompose the $A$-index into $A=\{0,a\} $ with $a=(1,\dots,D-1)$, and redefine the Poincar\'e generators according to
\begin{eqnarray}
P_0 &=& \omega H \,, \label{eq: P0 redef} \\
J_{0a} &=& \omega G_a \,, \label{eq: J0a redef}
\end{eqnarray}
where $H$ and $G_a$ are  the generators of time translations and boosts, respectively. The generators $P_a$ of space translations and $J_{ab}$ of spatial rotations are not redefined. Next, taking the limit $\omega \rightarrow \infty$ we obtain the following Carroll algebra:
\begin{eqnarray}\label{Calgebra}
&&[J_{ab}, P_c] = 2\delta_{c[a}P_{b]}\,,\hskip 1.5truecm [J_{ab}, G_c] = 2\delta_{c[a}G_{b]}\,,\nonumber\\[.2truecm]
&&[J_{ab}, J_{cd}] = 4
\delta_{[a[d}\,J_{c]b]}\,,\hskip 1.3truecm  [P_a,G_b] = \delta_{ab}H\,.
\end{eqnarray}
To each generator of the Carroll algebra we associate  a gauge field, a local parameter parametrizing
the corresponding symmetry and a  curvature, see Table \ref{Carrolltable}.

The gauge field transformations according to the Carroll algebra are given by
\begin{eqnarray} \label{bossymm1}
\delta \tau_\mu &=& \partial_\mu\zeta  -\omega_\mu{}^{a}\zeta^a+   e_\mu{}^a\lambda_a\,, \nonumber \\  [.1truecm]
\delta e_\mu{}^a &=& (D_\mu\zeta)^a +  \lambda^a{}_b e_\mu{}^b\,, \nonumber \\[.1truecm]
\delta \omega_\mu{}^{ab} &=&  (D_\mu\lambda)^{ab}\,,  \\[.1truecm]
\delta \omega_\mu{}^{a} &=& (D_\mu\lambda)^a +  \lambda^{a}{}_b \omega_{\mu }{}^b
\nonumber \,,
\end{eqnarray}
where $D_\mu$ is the covariant derivative with respect to spatial rotations, e.g., $(D_\mu\zeta)^a= \partial_\mu\zeta^a -\omega_\mu{}^{ab}\zeta^b$.
Like in the case of General Relativity, all gauge fields transform as covariant vectors under general coordinate transformations with parameter $\xi^\mu$.
In the following we will ignore the time and space translations but instead consider the general coordinate transformations.

{\small
\begin{table}[t]
\begin{center}
\begin{tabular}{|c|c|c|c|c|}
\hline
symmetry&generators& gauge field&parameters&curvatures\\[.1truecm]
\hline\rule[-1mm]{0mm}{6mm}
time translations&$H$&$\tau_\mu$&$\zeta(x^\nu)$&$R_{\mu\nu}(H)$\\[.1truecm]
space translations&$P_a$&$e_\mu{}^a$&$\zeta^a(x^\nu)$&$R_{\mu\nu}{}^a(P)$\\[.1truecm]
boosts&$G_a$&$\omega_\mu{}^{a}$&$\lambda^a(x^\nu)$&$R_{\mu\nu}{}^a(G)$\\[.1truecm]
spatial rotations&$J_{ab}$&$\omega_\mu{}^{ab}$&$\lambda^{ab}(x^\nu)$&$R_{\mu\nu}{}^{ab}(J)$\\[.1truecm]
\hline
\end{tabular}
\end{center}
\caption{This table indicates the generators of the Carroll algebra and the  gauge fields, local parameters and curvatures that are associated to each of these generators.}\label{Carrolltable}
\end{table}
 }

By construction the curvatures
\begin{align}\label{curvatures}
R_{\mu \nu}(H) &=  2 \partial_{[\mu} \tau_{\nu]} - 2\omega_{[\mu}{}^a e_{\nu] a}\,, \nonumber
\\[.1truecm]
R_{\mu \nu}{}^a(P) &=  2 \partial_{[\mu} e_{\nu]}{}^a - 2
\omega_{[\mu}{}^{ab} e_{\nu]b} \,, \nonumber \\[.1truecm]
R_{\mu\nu}{}^a(G) &= 2\partial_{[\mu}\omega_{\nu]}{}^{a} - 2
\omega_{[\mu}{}^{ab} \omega_{\nu]b}\,,
\\[.1truecm]
R_{\mu\nu}{}^{ab} (J) &=
2\partial_{[\mu}\omega_{\nu]}{}^{ab} - 2\omega_{[\mu}{}^a{}_c\,\omega_{\nu]}{}^{cb}  \,, \nonumber
\end{align}
transform covariantly under the Carroll transformations \eqref{bossymm1}.
In particular, they transform under Carroll boosts and spatial rotations as follows:
\begin{eqnarray}
\delta R_{\mu\nu}(H) &=& \lambda^aR_{\mu\nu}{}^a(P)  \,,\label{eq:varRH}\\[.1truecm]
\delta R_{\mu\nu}{}^a(P) &=&\lambda^{a b}R_{\mu\nu}{}^b(P)   \,,\label{eq:varRP} \\[.1truecm]
\delta R_{\mu\nu}{}^a(G) &=& \lambda^{a b}R_{\mu\nu}{}^b(G) - \lambda^b R_{\mu\nu}{}^{ab}(J) \,,\label{eq:varRG} \\[.1truecm]
\delta R_{\mu\nu}{}^{ab}(J)  &=&\lambda^{b c}R_{\mu\nu}{}^{ac}(J) - \lambda^{a c}R_{\mu\nu}{}^{bc}(J)\label{eq:varRJ} \,.
\end{eqnarray}
Furthermore, they satisfy the following Bianchi identities:
\begin{eqnarray}
\mathcal{D}_{[\mu} R_{\nu\rho]}{}(H) + R_{[\mu \nu}{}^{a}(G)e_{\rho]}^a &=& 0 \,,\label{eq: CarBianchi1}\\
\mathcal{D}_{[\mu} R_{\nu\rho]}{}^a(P) + R_{[\mu \nu}{}^{ab}(J)e_{\rho]}^b &=& 0 \,,\label{eq: CarBianchi2}\\
\mathcal{D}_{[\mu} R_{\nu\rho]}{}^a(G)  &=& 0 \,,\label{eq: CarBianchi3}\\
\mathcal{D}_{[\mu } R_{\nu \rho]}{}^{ab}(J)  &=& 0 \,, \label{eq: CarBianchi4}
\end{eqnarray}
where $\mathcal{D}_\mu$ is a Carroll-covariant derivative, i.e.~it is covariant with respect to Carroll boosts and spatial rotations.

\subsection{Carroll gravity}

We will first derive an invariant action for Carroll gravity by taking the ultra-relativistic limit of the action of General Relativity (\ref{eq: ActionS}). To define this limit, we redefine the gauge fields and symmetry parameters with the same parameter $\omega$ that occurs in the Carroll contraction defined by eqs.~\eqref{eq: P0 redef} and \eqref{eq: J0a redef}.
Requiring that the generalized parameter $\epsilon$ and generalized gauge field $A_\mu$ defined by
\begin{eqnarray}
\epsilon &=& \zeta H+\zeta^a P_a + \lambda^a G_a + \frac{1}{2} \lambda^{ab} J_{ab}\,,\label{generalized1}\\[.1truecm]
A_\mu &=& \tau_\mu H+e_\mu^a P_a + \omega_\mu^a G_a + \frac{1}{2} \omega_\mu^{ab} J_{ab} \,,\label{generalized2}
\end{eqnarray}
are invariant under the redefinitions leads to the following redefinitions of the gauge fields and parameters:
\begin{eqnarray}
E_\mu^0 &=& \omega^{-1} \tau_\mu \,, \quad \Omega^{0a}_\mu \;=\; \omega^{-1} \omega^a_\mu \,, \label{eq: rescal1}\\
E_\mu^a &=& e^a_\mu \,, \quad \quad \;\;\, \Omega^{ab}_\mu \;=\; \omega^{ab}_\mu \,, \label{eq: rescal2}\\
\eta^0 &=& \omega^{-1} \zeta \,, \quad \;\, \Lambda^{0a} \;=\; \omega^{-1} \lambda^a \,, \label{eq: rescal3}\\
\eta^a &=& \zeta^a \,, \quad \quad \;\;\, \Lambda^{ab} \;=\; \lambda^{ab} \,.\label{eq: rescal4}
\end{eqnarray}
One can show that performing these redefinitions in the relativistic transformation rules \eqref{Ptransf} and taking the limit $\omega \rightarrow \infty$ one recovers the Carroll transformations \eqref{bossymm1}.

Performing the same $\omega$-rescalings  (\ref{eq: rescal1}) and (\ref{eq: rescal2}) in the relativistic action (\ref{eq: ActionS}) we obtain
\begin{equation}\label{Carrollomega}
S_{{\tiny\mbox{Car}}}= - \frac{1}{16\pi G_N} \int \frac{e}{\omega} \left(2\tau^\mu e^\nu_a R(G)_{\mu\nu}{}^{a}+e^\mu_a e^\nu_b R(J)_{\mu\nu}{}^{ab} + O(\omega^{-2})\right)\,,
\end{equation}
where $e={\rm det}\, (\tau_\mu,e_\mu{}^a)$ is the ultra-relativistic determinant. We have defined here the projective inverses $\tau^\mu$ and $e^\mu{}_a$ according to:
\begin{alignat}{2} \label{inverses}
e_\mu{}^a e^\mu{}_b &= \delta^a_b \,, &\qquad \tau^\mu \tau_\mu = 1 \,, \nonumber \\[.2truecm]
\tau^\mu e_\mu{}^a &= 0 \,, & \tau_\mu e^\mu{}_a = 0 \,, \\[.2truecm]
e_\mu{}^a e^\nu{}_a &= \delta^\nu_\mu - \tau_\mu \tau^\nu\,.  & \nonumber
\end{alignat}
They transform under boosts and spatial rotations as follows:
\begin{eqnarray}
\delta \tau^\mu =0\,,\hskip 1.5truecm
\delta e^\mu_a  =  - \lambda^a\tau^\mu + \lambda^{ab}e_b^\mu \,. \label{eq: varinvviel Car}
\end{eqnarray}
Rescaling $G_N \rightarrow \omega^{-1} G_C$ and taking the $\omega\rightarrow\infty$ limit in the action \eqref{Carrollomega} we end up with the Carroll action\footnote{This limit shows similarities with the strong coupling limit considered in \cite{Henneaux:1979vn}, \cite{Henneaux:1981su},
see also \cite{Niedermaier:2014xwa}. Note that both limits lead to a theory with a Carroll-invariant vacuum solution. This suggests that, although looking different at first sight, the result of the two limits might be the same up to field redefinitions. We thank Marc Henneaux and Max Niedermaier for a discussion on this point.}
\begin{equation}
S_{{\tiny\mbox{Car}}}= - \frac{1}{16\pi G_C} \int e \left(2\tau^\mu e^\nu_a R(G)_{\mu\nu}{}^a+e^\mu_a e^\nu_b R(J)_{\mu\nu}{}^{ab}\right)\,. \label{eq: ActionCar}
\end{equation}
Using the variations \eqref{eq:varRG}, \eqref{eq:varRJ} and \eqref{eq: varinvviel Car} it can easily be checked that this action is invariant under Carroll boosts and rotations. In $D=3$, the Carroll algebra can be equipped with a non-degenerate, invariant bilinear form and as a consequence it is possible to write down a Chern-Simons action for the Carroll algebra. This Chern-Simons action is then equivalent to the one above.

The set of equations of motion obtained by varying $\tau^\mu$, $e^\mu_a$, $\omega_\mu^a$ and $\omega_\mu^{ab}$ in the Carroll action \eqref{eq: ActionCar}
can be written for any $D>2$ as follows:
\begin{eqnarray}
R_{\mu\nu}(H) &=&  0\,, \label{Careom1}\\
R_{\mu\nu}{}^a(P) &=& 0\,,\label{Careom4}\\
R_{\mu a}{}^a(G) &=& 0 \,,\label{Careom5}\\
R_{0b}{}^{ab}(J) &=& 0 \,,\\
R_{ac}{}^{bc}(J) + R_{0a}{}^b(G) &=& 0 \,,\label{Careom7}
\end{eqnarray}
where $R_{0b}{}^{ab}(J)=\tau^\mu e^\nu_b R_{\mu\nu}{}^{ab}(J)$ and we are using the same notation for the remaining projections of the curvatures. The equations (\ref{Careom1}) - (\ref{Careom4}) can be used to solve for the spin connections
\begin{eqnarray}
\omega_{\mu}{}^a &=& \tau_\mu \tau^\nu e^{\rho a} \partial_{[\nu}\tau_{\rho]} + e^{\nu a}\partial_{[\mu}\tau_{\nu]} + S^{ab} e^b_\mu \,, \label{eq: solomegaa}\\[.1truecm]
\omega_\mu{}^{ab} &=& - 2 e^{\rho [a}\partial_{[\mu} e_{\rho]}^{b]} + e_{\mu c} e^{\rho a} e^{\nu b}\partial_{[\rho} e_{\nu]}^c  \,,\label{eq: solomegaab}
\end{eqnarray}
except for a symmetric component $S^{ab}=S^{(ab)} = e^{\mu(a} \omega_\mu^{b)}$ of the boost spin connection $\omega_\mu{}^a$ which remains undetermined. Below we will give an interpretation for $S^{ab}$. The equation \eqref{Careom4} can additionally be used to derive the constraint
\begin{equation}
K_{ab} = 0  \,, \label{eq: constraint viel}
\end{equation}
where we defined $K_{ab} = e_a^\mu e_b^\nu  K_{\mu\nu}$ with $K_{\mu\nu}$ the extrinsic curvature given by the Lie derivative of $h_{\mu\nu}=e_\mu^a e_\nu^b \delta_{ab}$ along the vector field $\tau^\mu$
\begin{equation}
K_{\mu\nu} \equiv \frac{1}{2}\mathcal{L}_{\tau}(h_{\mu\nu}) = \frac{1}{2}\left(\tau^\rho\partial_\rho h_{\mu\nu}+ h_{\mu\rho}\partial_\nu \tau^\rho + h_{\nu\rho}\partial_\mu \tau^\rho \right)\,. \label{eq: ExtrinsicCurLie}
\end{equation}
The fact that curvature constraints are not only used to solve for (part of) the spin-connections but also lead to constraints on the geometry has been encountered before in the construction of the so-called stringy Newton-Cartan gravity theory \cite{Andringa:2012uz}.

Let us stress that from equations \eqref{eq: solomegaa} and \eqref{eq: solomegaab} by themselves, it follows that the spin connections transform under Carroll boosts and rotations according to
\begin{eqnarray}
\delta \omega^{a}_\mu &=& (D_\mu\lambda)^a +  \lambda^{a}{}_b \omega_{\mu }{}^b + \tau_\mu \delta^{ab}K_{bc}\lambda^c \,, \label{eq: var sol omegaa}\\
\delta \omega^{ab}_\mu &=& D_\mu \lambda^{ab} - 2\lambda^{[a} \delta^{b]c} e_{\mu}^d K_{cd} \,. \label{eq: var sol omegaab}
\end{eqnarray}
Hence, it is only thanks to the constraint \eqref{eq: constraint viel} imposed on the geometry that the transformation of the spin connections agrees with \eqref{bossymm1}. In order to obtain \eqref{eq: var sol omegaa} we used
\begin{equation}
\delta S^{ab} = \lambda^{ac}S^{bc}+\lambda^{bc}S^{ac} + e^{\mu (a}\partial_\mu\lambda^{b)} - \lambda^{(a}\omega^{b)}_\mu\tau^\mu
- \lambda^{c}e^{\nu (a}\omega^{b)c}_{\nu}  \,, \label{eq: varS}
\end{equation}
as can be directly deduced from \eqref{bossymm1} since $S^{ab} = e^{\mu (a} \omega^{b)}_\mu$. 

The geometrical constraint \eqref{eq: constraint viel} is closely related to the undetermined components $S^{ab}$ of the boost spin connection. In order to see this, it is instructive to go to a second order formulation of Carroll gravity. Plugging the dependent expressions for the spin connections \eqref{eq: solomegaa} and \eqref{eq: solomegaab} into the Carroll action \eqref{eq: ActionCar} we obtain
\begin{equation} \label{eq: Car act S lag multi}
S_{{\tiny\mbox{Car}}}= - \frac{1}{16\pi G_C} \int e \left( \left.2\tau^\mu e^\nu_a R(G)^a{}_{\mu\nu}\right|_{S^{ab}=0}+e^\mu_a e^\nu_b R(J)^{ab}{}_{\mu\nu} + 2K_{ab}S^{ab}-2\delta^{ab}\delta_{cd}K_{ab}S^{cd}\right)\,, \end{equation}
where we performed an integration by part on the $S^{ab}$ dependent terms.\footnote{This implies that we end up with a Lagrangian that is only Carroll invariant up to total derivative terms.} From the expression \eqref{eq: Car act S lag multi} for the action it follows that the equation of motion for $S^{ab}$ implies $K_{ab}=0$. In other words, we conclude that the $S^{ab}$ term is actually a Lagrange multiplier that enforces the constraint \eqref{eq: constraint viel} which, previously in the first order formulation, was a consequence of the equations of motion for the spin connections.

Finally, Carroll gravity can be rewritten in a second order metric formulation in terms of the fields $\tau_\mu$ , $h_{\mu\nu}$ and $S^{\mu\nu}=e^\mu_ae^\nu_bS^{ab}$. In order to do this we first trade the spin connections for a Christoffel connection. The spin connections can be related to a space-time connection by imposing a vielbein postulate
\begin{eqnarray}
\partial_\mu \tau_\nu - \Gamma^\rho_{\mu\nu}\tau_\rho - \omega^a_\mu e_\nu^a  &\equiv& 0 \,, \label{eq: vp1} \\
\partial_\mu e^a_\nu - \Gamma^\rho_{\mu\nu}e^a_\rho - \omega^{ab}_\mu e_\nu^b &\equiv& 0\,.\label{eq: vp2}
\end{eqnarray}
The vielbein postulate implies the following relation between the space-time connection $\Gamma^\rho_{\mu\nu}$ and the spin connections
\begin{equation}
\Gamma^{\rho}_{\mu \nu} = \tau^{\rho}\partial_{\mu}{\tau_{\nu}} + e_{a}^{\rho}\partial_{\mu}{e^{a}_{\nu}} - \tau^{\rho}\omega^{a}_{\mu}e^{b}_{\nu}\delta_{a b} - e_{a}^{\rho}\omega^{a b}_{\mu}e^{c}_{\nu}\delta_{b c} \,. \label{eq: Gammacon VP}
\end{equation}
A few remarks are in order here. By construction, the connection $\Gamma^{\rho}_{\mu \nu}$ would be Carroll invariant if the fields would transform as in \eqref{bossymm1}. However, this is not the case at this stage since we have additional $K_{ab}$ contributions in \eqref{eq: var sol omegaa} and \eqref{eq: var sol omegaab}. Also, on general grounds it follows from the vielbein postulate that
\begin{equation}
K_{\mu\nu} =  \left( {h}_{\mu \rho}{\Gamma}^{\rho}_{[\sigma \nu]}  +{h}_{\nu \rho}{\Gamma}^{\rho}_{[\sigma \mu]} \right)\tau^\sigma  \,,
\end{equation}
where $\Gamma^{\rho}_{[\mu\nu]}$ represents the torsion. Hence, on a Carrollian geometry $K_{\mu\nu}$ is automatically vanishing whenever there is no spatial component to the torsion, namely whenever $e^a_\rho\Gamma^{\rho}_{[\mu\nu]}$ vanishes which is precisely the content of equation \eqref{Careom4}. The same constraint on the torsion also occurs in the context of the Carroll geometry of \cite{Hofman:2014loa}.

Now let us rewrite $\Gamma^{\rho}_{\mu \nu}$ in a metric formulation. Plugging \eqref{eq: solomegaa} and \eqref{eq: solomegaab} into \eqref{eq: Gammacon VP} we obtain
\begin{eqnarray}
\Gamma^{\rho}_{\mu \nu} &=& \tau^{\rho}\left(\partial_{(\mu}{\tau_{\nu)}} + \tau_{\mu}\tau^{\sigma}\partial_{[\nu}{\tau_{\sigma]}}
+ \tau_{\nu}\tau^{\sigma}\partial_{[\mu}{\tau_{\sigma]}}  - h_{\mu\tau}h_{\nu\sigma}S^{\tau\sigma} \right) \nonumber\\&&
- h^{\rho \sigma}K_{\sigma \mu}\tau_{\nu}  + \frac{1}{2}h^{\rho \sigma}\left(\partial_{\mu}{h_{\nu \sigma}} + \partial_{\nu}{h_{\mu \sigma}} - \partial_{\sigma}{h_{\mu \nu}}\right) \,. \label{eq: Gammacon expli}
\end{eqnarray}
We then define a Riemann tensor with respect to the connection $\Gamma^{\rho}_{\mu \nu}$ in the usual way
\begin{eqnarray}
R_{\mu\nu\rho}{}^{\sigma} &=& - \partial_{\mu}{\Gamma^{\sigma}_{\nu \rho}} + \partial_{\nu}{\Gamma^{\sigma}_{\mu \rho}} - \Gamma^{\sigma}_{\mu \lambda}\Gamma^{\lambda}_{\nu \rho} + \Gamma^{\sigma}_{\nu \lambda}\Gamma^{\lambda}_{\mu \rho} \,. \label{eq: Riemann def} 
\end{eqnarray}
Finally, the Carroll invariant action in a second order metric formulation reads \footnote{
Alternatively, we can define a modified connection $\hat \Gamma^\rho_{\mu\nu} = \Gamma^\rho_{\mu\nu}  - \tau^\rho \partial_{[\mu}\tau_{\nu]}  +  \tau^\rho\tau^\sigma\left(\tau_\mu\partial_{[\nu}\tau_{\sigma]} +  \tau_\nu \partial_{[\mu}\tau_{\sigma]}\right)  - \tau^\rho S_{\mu\nu}$ such that the action takes the simpler form $S_{{\tiny\mbox{Car}}}= \frac{1}{16\pi G_C} \int e h^{\mu \nu} \hat R_{\mu \nu}$ with $\hat R_{\mu \nu}$ the Ricci tensor relative to the shifted connection $\hat \Gamma^\rho_{\mu\nu}$.}
\begin{equation}
S_{{\tiny\mbox{Car}}}= \frac{1}{16\pi G_C} \int e h^{\mu \nu}\left( R_{\mu \nu} \frac{}{}+ \tau^\rho\tau_\sigma R_{\mu\rho\nu}{}^\sigma \right) \,, \label{eq: car action sec}
\end{equation}
with $\Gamma^\rho_{\mu\nu}$ given by \eqref{eq: Gammacon expli} and where we defined the Ricci tensor as $R_{\mu\nu}=R_{\mu\sigma\nu}{}^{\sigma}$. Since we have seen that in the second order formulation the connection $\Gamma^\rho_{\mu\nu}$ is not Carroll invariant $\delta\Gamma^\rho_{\mu\nu}\neq0$, it follows that the invariance of the action \eqref{eq: car action sec} is no longer manifest. 

In the second order formulation, the equations of motion for $S^{\mu\nu}$ read
\begin{equation}
K_{\mu \nu} - h_{\mu\nu}K = 0\,,
\end{equation}
with $K=h^{\mu\nu}K_{\mu\nu}$ and for $D>2$ this implies that $K_{\mu \nu}=0$. We thus reproduce the constraint we initially obtained in the first order formalism. As we already learned from equation \eqref{eq: Car act S lag multi} $S^{\mu\nu}$ is hence to be seen as a Lagrange multiplier whose role is to impose this constraint on the geometry. Using that $K_{\mu\nu}=0$ the remaining equations of motion obtained by varying $\tau^\mu$ and $h^{\mu\nu}$ \footnote{In varying $h^{\mu\nu}$ one should use that its variation is constrained due to $h^{\mu\nu} \tau_\nu = 0$. This implies that one should take care of projecting out the purely time-like components of the equation obtained by varying $h^{\mu\nu}$. E.g., upon varying $h^{\mu\nu} X_{\mu\nu}$, where $X_{\mu\nu}$ does not depend on $h^{\mu\nu}$, the correct equation of motion is $X_{\mu\nu} - \tau_\mu \tau_\nu \tau^{\rho} \tau^{\sigma} X_{\rho \sigma} = 0$.} are
\begin{eqnarray}
\left(\tau_\lambda h_\mu{}^{\sigma} - \frac{1}{2}\tau_{\mu}h_{\lambda}{}^{\sigma}\right)h^{\nu\rho}R_{\sigma\nu\rho}{}^{\lambda} &=& 0 \,,\\
R_{\mu\nu} - \frac{1}{2}h_{\mu\nu}\hat R &=&  0\,,\label{eq: eom 2nd Car}
\end{eqnarray}
with $h_\mu{}^\nu=h_{\mu\rho}h^{\nu\rho}$ and $\hat R = h^{\mu \nu}R_{\mu \nu}+ \tau^\rho\tau_\sigma h^{\mu \nu}R_{\mu\rho\nu}{}^\sigma$. Note that with $K_{\mu\nu}=0$ the terms $h_{\mu\nu}$, $R_{\mu\nu}$ and $\hat R$ in equation \eqref{eq: eom 2nd Car} are all separately Carroll invariant. Moreover, in this case, the Ricci tensor becomes symmetric and since it satisfies $R_{\mu\nu}\tau^\nu=0$ equation \eqref{eq: eom 2nd Car} leads to $\frac{1}{2}D(D-1)$ equations.

The Carroll theory we described in this section can be compared to the Carroll geometry developed in \cite{Hartong:2015xda}. In  \cite{Hartong:2015xda} the extrinsic curvature $K_{\mu\nu}$ is not constrained to vanish but is kept arbitrary. Moreover, in \cite{Hartong:2015xda} the Carroll symmetries are realised on the fields $\tau_\mu, h_{\mu\nu}$ and a vector field $M^\mu=e^\mu_a M^a$. This is different from the present case where the additional field needed to realise the Carroll symmetries is a symmetric tensor $S^{ab}$. Furthermore, although when evaluated in the case $K_{\mu\nu}=0$ the rotation spin connection \eqref{eq: solomegaab} agrees precisely with the one obtained in \cite{Hartong:2015xda}, there exists no special choice of $S^{ab}$ such that the boost spin connection \eqref{eq: solomegaa} would match the one of \cite{Hartong:2015xda}. The reason for this is that in the latter case the boost connection is by construction always of the form
\begin{equation}
\mbox{boost connection of}\; \cite{Hartong:2015xda}: \qquad \omega^a_{\mu} = \partial_\mu M^a -\omega^{ab}_\mu M^b\,.
\end{equation}
In particular, $\tau^\mu\omega^a_{\mu}$ is then a function of $M^a$ whereas in our case $\tau^\mu\omega^a_{\mu}$ is not a function of $S^{ab}$. Hence, there cannot be a choice of $S^{ab}$ for which the connections would agree. For further comments, see the conclusions.

\section{Galilei Gravity}

The kinematics of Galilei gravity can be obtained by gauging the Galilei algebra. In contrast to Newton-Cartan gravity, Galilei gravity has no mass parameter.
In this section we will perform the same steps as for Carroll gravity thereby emphasizing the similarities as well as the differences.  In the first subsection we will review a few
properties of the Galilei algebra while in the second subsection we will construct Galilei gravity.

\subsection{The Galilei Algebra}

The Galilei algebra is obtained by a contraction of the Poincar\'e algebra. To define this contraction, we decompose the $A$-index into $A=\{0,a\} $ with $a=(1,\dots,D-1)$, and redefine the Poincar\'e generators according to
\begin{eqnarray}
P_0 &=& \omega^{-1} H \,, \label{eq: GalP0 redef} \\
J_{0a} &=& \omega G_a \,, \label{eq: GalJ0a redef}
\end{eqnarray}
where $H$ and $G_a$ are  the generators of time translations and boosts, respectively. The generators $P_a$ of space translations and $J_{ab}$ of spatial rotations are not redefined. Next, taking the limit $\omega \rightarrow \infty$ we obtain the following Galilei algebra:
\begin{eqnarray}\label{Galgebra}
&&[J_{ab}, P_c] = 2\delta_{c[a}P_{b]}\,,\hskip 1.5truecm [J_{ab}, G_c] = 2\delta_{c[a}G_{b]}\,,\nonumber\\[.2truecm]
&&[J_{ab}, J_{cd}] = 4
\delta_{[a[d}\,J_{c]b]}\,,\hskip 1.3truecm  [H,G_a] = P_a\,.
\end{eqnarray}
To each generator of the Galilei algebra we associate  a gauge field, a local parameter parametrizing
the corresponding symmetry and a  curvature, for which we use the same notation as in the case of the Carroll algebra, see Table \ref{Carrolltable}.

The gauge field transformations according to the Galilei algebra are given by
\begin{eqnarray} \label{Gtransf1}
\delta \tau_\mu &=& 0\,,  \\[.1truecm]
\delta e_\mu{}^a  &=& \lambda^a\tau_\mu + \lambda^{ab} e_\mu{}^b  \,, \\[.1truecm]
\delta \omega_\mu{}^{ab}  &=& (D_\mu \lambda)^{a b}, \label{eq: Gal trans omegaab}\\[.1truecm]
\delta \omega_\mu{}^a  &=& (D_\mu \lambda)^a + \lambda^{a}{}_b\omega^b_\mu  \,.\label{Gtransf4}
\end{eqnarray}
Like in the Carroll case, all gauge fields transform as covariant vectors under general coordinate transformations with parameter $\xi^\mu$.
In the following we will ignore the time and space translations but instead consider the general coordinate transformations.

The curvatures that transform covariantly under the Galilei transformations \eqref{Gtransf1}-\eqref{Gtransf4} are given by
\begin{eqnarray}
R_{\mu\nu}(H) &=& 2\partial_{[\mu}\tau_{\nu]}  \,,\label{eq:GalRH}\\
R_{\mu\nu}{}^a(P) &=& 2\partial_{[\mu}e^a_{\nu]} - 2\omega_{[\mu}^{ab} e^b_{\nu]} - 2\omega_{[\mu}^{a} \tau_{\nu]}\,.\label{eq:GalRP} \\
R_{\mu\nu}{}^a(G) &=& 2\partial_{[\mu} \omega^a_{\nu]}  -  2\omega^{ab}_{[\mu} \omega^b_{\nu]} \,, \\[.1truecm]
R_{\mu\nu}{}^{ab}(J)  &=& 2\partial_{[\mu} \omega^{ab}_{\nu]}  - 2 \omega^{ac}_{[\mu} \omega^{cb}_{\nu]}  \,.
\end{eqnarray}
They transform under Galilean boosts and spatial rotations as follows:
\begin{eqnarray}
\delta R_{\mu\nu}(H) &=& 0 \,,\label{eq:GalvarRH}\\[.1truecm]
\delta R_{\mu\nu}{}^a(P) &=&\lambda^{a b}R_{\mu\nu}{}^b(P)  + \lambda^a R_{\mu\nu}(H) \,.\label{eq:GalvarRP} \\[.1truecm]
\delta R_{\mu\nu}{}^a(G) &=& \lambda^{a b}R_{\mu\nu}{}^b(G) - \lambda^b R_{\mu\nu}{}^{ab}(J) \,, \\[.1truecm]
\delta R_{\mu\nu}{}^{ab}(J)  &=&\lambda^{b c}R_{\mu\nu}{}^{ac}(J) - \lambda^{a c}R_{\mu\nu}{}^{bc}(J)
\end{eqnarray}
and satisfy the following Bianchi identities:
\begin{eqnarray}
\mathcal{D}_{[\mu} R_{\nu\rho]}{}(H) &=& 0 \,,\label{eq: GalBianchi1}\\
\mathcal{D}_{[\mu} R_{\nu\rho]}{}^a(P) + R_{[\mu \nu}{}^{a}(G)\tau_{\rho]} + R_{[\mu \nu}{}^{ab}(J)e_{\rho]}^b &=& 0 \,,\label{eq: GalBianchi2}\\
\mathcal{D}_{[\mu} R_{\nu\rho]}{}^a(G)  &=& 0 \,,\label{eq: GalBianchi3}\\
\mathcal{D}_{[\mu } R_{\nu \rho]}{}^{ab}(J)  &=& 0 \,, \label{eq: GalBianchi4}
\end{eqnarray}
where $\mathcal{D}_\mu$ is a Galilei-covariant derivative, i.e.~it is covariant with respect to Galilei boosts and spatial rotations.

\subsection{Galilei gravity}

Like in the Carroll case  an invariant action for Galilei  gravity  can be obtained by taking the non-relativistic limit of the action of General Relativity (\ref{eq: ActionS}). To define this limit we redefine the gauge fields and symmetry parameters with the same parameter $\omega$ that occurs in the Carroll contraction defined by eqs.~\eqref{eq: GalP0 redef} and \eqref{eq: GalJ0a redef}.
Requiring that the generalized parameter $\epsilon$ and generalized gauge field $A_\mu$ defined by eqs.~\eqref{generalized1} and \eqref{generalized2}
are invariant under the redefinitions leads to the following redefinitions of the gauge fields and parameters:
\begin{eqnarray}
E_\mu^0 &=& \omega \tau_\mu \,, \;\;\quad \Omega^{0a}_\mu \;=\; \omega^{-1} \omega^a_\mu \,, \label{eq: Galrescal1}\\
E_\mu^a &=& e^a_\mu \,, \quad \;\;\;\, \; \Omega^{ab}_\mu \;=\; \omega^{ab}_\mu \,, \label{eq: Galrescal2}\\
\eta^0 &=& \omega \zeta \,, \;\;\quad \;\, \Lambda^{0a} \;=\; \omega^{-1} \lambda^a \,, \label{eq: Galrescal3}\\
\eta^a &=& \zeta^a \,, \quad \;\; \;\;\, \Lambda^{ab} \;=\; \lambda^{ab} \,.\label{eq: Galrescal4}
\end{eqnarray}

Performing the same $\omega$-rescalings (\ref{eq: Galrescal1}) and (\ref{eq: Galrescal2}) in the relativistic action (\ref{eq: ActionS}),
rescaling $G_N \rightarrow \omega G_G$ and taking the $\omega\rightarrow\infty$ limit  we end up with the following Galilei  action
\begin{equation}
S_{{\tiny\mbox{Gal}}}= - \frac{1}{2\kappa} \int e  R_{\mu\nu}{}^{ab}(J) e^\mu_a e^\nu_b\,,\label{eq: ActionGal}
\end{equation}
where $\kappa = 8\pi G_G$ and $e={\rm det}\, (\tau_\mu,e_\mu{}^a)$ is the non-relativistic determinant. We have used here the same definition of the projective inverses
$\tau^\mu$ and $e^\mu{}_a$ like in the Carroll case, see eq.~\eqref{inverses}. These projective inverses transform under the Galilei boosts and spatial rotations as follows:
\begin{eqnarray}
\delta \tau^\mu &=& - \lambda^a e_a^\mu \,, \hskip 2truecm
\delta e^\mu_a  =  \lambda^{ab}e_b^\mu \,.
\end{eqnarray}
One may verify that the Galilei action \eqref{eq: ActionGal} is not only Galilei invariant but it also has an accidental local scaling symmetry given by
\begin{eqnarray}
\tau_\mu & \rightarrow & \lambda(x)^{-(D-3)} \tau_\mu  \,,\label{eq: rescaltau}\\
e^a_\mu &\rightarrow & \lambda(x) e^a_\mu \,,\label{eq: rescale}
\end{eqnarray}
where $\lambda(x)$ is an arbitrary function. Hence, the full invariance of the Galilean gravity action is that of a Schr\"odinger algebra without central charge and with critical exponent $z=-(D-3)$.

For any $D>2$ the equations of motion that follow from the variation of the Galilei action \eqref{eq: ActionGal} are 
equivalent to a constraint on the geometry
\begin{equation}
R_{ab}(H) = e^\mu_a e^\nu_b \partial_{[\mu}\tau_{\nu]} =  0\,, \label{eq: RHab is zero}
\end{equation}
together with the following equations
\begin{eqnarray}
R_{0a}(H) &=&  \frac{D-3}{D-2} R_{ab}{}^{b}(P)\,, \label{eq: Gal eom1}\\[.1truecm]
R_{ab}{}^{c}(P) &=& -\frac{2}{D-2} \delta_{[a}^c R_{b]d}{}^{d}(P)\,,\label{eq: Gal eom2}\\[.1truecm]
R_{\mu b}{}^{ab}(J) &=& 0 \,.
\end{eqnarray}
The constraint \eqref{eq: RHab is zero} means that this geometry has twistless torsion \cite{Christensen:2013rfa}. Clearly, we see from \eqref{eq: Gal eom1} that $D=3$ is special, we will come back to this case below and first assume $D>3$.

For $D>3$ the equation of motion \eqref{eq: Gal eom1} and \eqref{eq: Gal eom2} can be used to solve for the spatial rotation spin connection $\omega_\mu{}^{ab}$ as
\begin{equation}
\omega_\mu^{ab} = \tau_{\mu}A^{a b} + e_{\mu c} \left(e^{\rho [a} e^{b]\nu}\partial_{\rho}{e^c_{\nu}} + e^{\rho[a} e^{c]\nu}\partial_{\rho}{e^{b}_{\nu}} - e^{\rho[b} e^{c]\nu}\partial_{\rho}{e^{a}_{\nu}}\right)
+ \frac{4}{D-3} e^{\rho[a} e^{b]}_{\mu} \tau^{\nu}\partial_{[\rho}\tau_{\nu]}\,, \label{eq: Gal solomegaab}
\end{equation}
except for $A^{ab}$ which is an undetermined anti-symmetric tensor component of $\omega_\mu{}^{ab}$.

The constraint \eqref{eq: RHab is zero} is a restriction on the geometry which can be seen as the Galilean equivalent to the constraint \eqref{eq: constraint viel} in the Carroll case. In the second order formulation the constraint \eqref{eq: RHab is zero} arises from the variation with respect to $A^{ab}$. Hence, we can interpret $A^{ab}$ as a Lagrange multiplier. Indeed, in the case $D>3$, plugging \eqref{eq: Gal solomegaab} into the action \eqref{eq: ActionGal} to obtain it in a second order formulation leads to
\begin{equation}
S_{{\tiny\mbox{Gal}}}= - \frac{1}{2\kappa} \int e \left( \left. R_{\mu\nu}{}^{ab}(J) e^\mu_a e^\nu_b \right|_{A^{ab}=0} + A^{ab}R_{ab}(H) \right) \,.\label{eq: ActionGal sec order}
\end{equation}
This makes manifest the fact that the variation with respect to $A^{ab}$ of the second order action in equation \eqref{eq: ActionGal sec order} reproduces the constraint \eqref{eq: RHab is zero}.

The field $A^{ab}$ does not transform covariantly, as can be seen from \eqref{eq: Gal trans omegaab}. Since $A^{ab}$ is undetermined we can make a redefinition
\begin{equation}
\bar A^{ab} = A^{ab} + \tau^{\rho}e^{\mu [a}\partial_{\rho}e^{b]}_\mu \,, \label{eq: redef barA}
\end{equation}
such that $\bar A^{ab}$ transforms covariantly
\begin{equation}
\delta \bar A^{ab} = \lambda^{ac}\bar A^{cb} + \lambda^{bc}\bar A^{ac}  - \lambda^c e_c^\mu \omega^{ab}_\mu - \lambda^c e_c^\mu e^{\nu [a} \partial_\mu e^{b]}_\nu - \lambda^{[a}e_c^{b]\nu} \tau^\mu  \partial_\mu \tau_\nu \,. \label{eq: trans barAab}
\end{equation}
The solution for $\omega^{ab}_\mu$ given in equation \eqref{eq: Gal solomegaab} transforms according to
\begin{equation}
\delta \omega_\mu^{ab} = (D_\mu \lambda){}^{ab} - \left(  2\lambda^{[a} {e}^{b]\nu} {e}_{c}^{\rho}  -  e^{a\nu} e^{b\rho}\lambda^c\right){e}^{c}_{\mu} \partial_{[\nu} \tau_{\rho]}  + \frac{4}{D-3}e^{[a}_{\mu} e^{b]\nu} {\lambda}^c {e}_{c}^{\rho}  \partial_{[\nu} \tau_{\rho]}\,. \label{eq: var sol omegaab gal}
\end{equation}
Similar to the Carroll case, this transformation agrees with \eqref{eq: Gal trans omegaab} only up to the geometrical constraint $e^\mu_a e^\nu_b \partial_{[\mu}\tau_{\nu]} =0$ which we found in equation \eqref{eq: RHab is zero}.

We will now rewrite the action \eqref{eq: ActionGal sec order} in a second order metric formulation in terms of $\tau_\mu$, $h_{\mu\nu}$ and $\bar A^{\mu\nu}$ following the same steps as we did in the Carroll case. This time however it will be necessary to use the redefined $\bar A^{\mu\nu}= e^\mu_a e^\nu_b \bar A^{ab}$ of equation \eqref{eq: redef barA} instead of $A^{\mu\nu}= e^\mu_a e^\nu_b A^{ab}$ in order to fully remove all vielbeins $e^a_\mu$ and obtain the theory in a metric formulation. Proceeding in a similar manner as before, namely trading the spin connections for a Christoffel connection $\Gamma^\rho_{\mu\nu}$ by imposing a vielbein postulate, we obtain\,\footnote{A related action occurs in  \cite{DePietri:1994je} as the leading term in the
non-relativistic expansion of an ADM formulation of the Einstein-Hilbert Lagrangian. This work does not mention, however, the occurrence of Galilean symmetries in this leading term.}
\begin{equation}
S_{{\tiny\mbox{Gal}}}= \frac{1}{2\kappa} \int e h^{\mu\nu}R_{\mu\nu} \,,\label{eq: ActionGal sec order metric}
\end{equation}
with the same definitions for the Riemann and Ricci tensors we used before, see equation \eqref{eq: Riemann def} and below equation \eqref{eq: car action sec}. In this case the $\Gamma^\rho_{\mu\nu}$ connection that follows from the vielbein postulate and appears in equation \eqref{eq: ActionGal sec order metric} is given by
\begin{eqnarray}
\Gamma^\rho_{\mu\nu} &=& \tau^\rho \partial_\mu \tau_\nu + e_a^\rho \partial_\mu e^a_\nu - e^\rho_a\omega_\mu{}^a \tau_\nu - e_a^\rho \omega^{ab}_\mu e^b_\nu \,, \label{eq: Gal Gamma VP}\\ 
&=& \tau^{\rho}\partial_{\mu}{\tau_{\nu}} - e_{a}^{\rho}\omega^{a}_{\mu}\tau_{\nu}
+ \frac{2}{D-3}h^{\rho \sigma}\tau^{\lambda}\left(h_{\mu\nu}\partial_{[\lambda}{\tau_{\sigma]}} + h_{\mu \sigma}\partial_{[\nu}{\tau_{\lambda]}}\right) - \tau_{\mu}h_{\nu \sigma}\bar A^{\rho\sigma}  - h^{\rho \sigma} K_{\sigma \mu}\tau_{\nu}  \nonumber\\&&
 + \frac{1}{2}h^{\rho \sigma}\left(\partial_{\mu}{h_{\nu \sigma}} + \partial_{\nu}{h_{\mu \sigma}} - \partial_{\sigma}{h_{\mu \nu}}\right) + \frac{1}{2}\tau_\mu h_\nu{}^{\tau}h^{\rho \sigma} \tau^\lambda   \left( \partial_\sigma h_{\tau\lambda} -  \partial_\tau h_{\sigma\lambda}\right)\,. \qquad \label{eq: Gal Gamma}
\end{eqnarray}
A few remarks are in order. First of all, note that due to the fact that $\Gamma^\rho_{\mu\nu}$ is obtained directly from a vielbein postulate, equation \eqref{eq: Gal Gamma VP} being the result, the boost spin connection $\omega^a_\mu$ naturally appears in $\Gamma^\rho_{\mu\nu}$. However, as expected all the terms containing $\omega^a_\mu$ automatically cancel out in the action \eqref{eq: ActionGal sec order metric}, leaving us with a second order formulation for Galilei gravity that depends only on $\tau_\mu$, $h_{\mu\nu}$ and $\bar A^{\mu\nu}$. Here, the use of $\bar A^{\mu\nu}$ over $A^{\mu\nu}$ is necessary since the difference between these two terms cannot be rewritten without using the vielbein $e^a_\mu$, see equation \eqref{eq: redef barA}. In the second order formulation, the connection $\Gamma^\rho_{\mu\nu}$ is not  Galilean invariant. This is due to the fact that the spin connection $\omega^{ab}_\mu$ which appears in \eqref{eq: Gal Gamma VP} transforms according to \eqref{eq: var sol omegaab gal} instead of \eqref{eq: Gal trans omegaab}. As a direct consequence of this, the Lagrangian given in equation \eqref{eq: ActionGal sec order metric} is not an invariant. However, as we already observed in the Carroll case, the action is invariant.

The equations of motion obtained by varying the Galilean action \eqref{eq: ActionGal sec order metric} with respect to $\bar A^{\mu\nu}$ are
\begin{equation}
{h}_{\mu}{}^{\rho} {h}_{\nu}{}^{\sigma} \left(\partial_\rho\tau_\sigma - \partial_\sigma \tau_\rho\right) = 0 \,. \label{eq: gal constraint metric}
\end{equation}
As expected this is nothing else than the constraint \eqref{eq: RHab is zero}. Using this geometric constraint, the remaining equations of motion obtained by varying the action with respect to $\tau^\mu$ and $h^{\mu\nu}$, respectively, read
\begin{eqnarray}
\tau_\mu R &=& 0 \,, \\
h^{\rho\sigma}h_{\sigma(\mu} R_{\nu)\rho}   - \frac{1}{2} h_{\mu \nu} R - \tau_{(\mu}  h_{\nu) \tau} \tau^{\rho}h^{\sigma \lambda} R_{\rho \sigma \lambda}{}^{\tau}  &=& 0 \,,
\end{eqnarray}
with $R=h^{\mu\nu}R_{\mu\nu}$. Note that in this case $R_{\mu\nu}$ is not symmetric but both the Ricci and the Riemann tensors become invariants whenever the constraint \eqref{eq: gal constraint metric} is satisfied.

The case $D=3$ is special. In that case we may write $\omega_\mu{}^{ab} =\epsilon^{ab}\omega_\mu$ and it can be seen from the first order equations of motion \eqref{eq: RHab is zero}-\eqref{eq: Gal eom2} that the whole $\omega_\mu$ remains undetermined. Hence, an interesting consequence is that there is intrinsically no second order formulation for $D=3$. Also, in contrast to the $D>3$ case, the equations of motion imply a stronger geometrical constraint, namely
\begin{equation}
R_{\mu\nu}(H) = \partial_{[\mu}\tau_{\nu]} = 0 \,. \label{eq: RH0a is zero}
\end{equation}
Using the identity $e\epsilon^{ab}e_a^\mu e_b^\nu = 2\epsilon^{\mu\nu\rho}\tau_\rho$, which is valid for $D=3$, the Galilean action \eqref{eq: ActionGal} can be rewritten as
\begin{equation}
S_{{\tiny\mbox{Gal 3D}}}= - \frac{1}{2\kappa} \int \epsilon^{\mu\nu\rho} \tau_\mu\partial_\nu\omega_\rho\,.\label{eq: ActionGal 3D}
\end{equation}
This form of the action makes manifest that its variation with respect to $\omega_\mu$ precisely reproduces the constraint obtained in equation \eqref{eq: RH0a is zero}. Note that the Galilei algebra in $D=3$ only allows for a degenerate invariant bilinear form. The above action corresponds to the Chern-Simons action for the Galilei algebra with this degenerate bilinear form. The degeneracy of the form explains why not all fields occur in the action.

\section{Matter Couplings}

We generalize the discussion so far to include  matter couplings. For this purpose, we consider the action
\begin{equation}\label{total}
S_{{\tiny\mbox{Tot}}} = S_{{\tiny\mbox{grav}}} + S_{{\tiny\mbox{mat}}}\,,
\end{equation}
where $S_{{\tiny\mbox{grav}}}$ will be either Carroll or Galilei gravity and $S_{{\tiny\mbox{mat}}}$ denotes a general matter action. We define the following currents
\begin{eqnarray}
J^\mu{}_a &=& \frac{\kappa}{e}\frac{\delta S_{{\tiny\mbox{mat}}}}{\delta \omega^a_\mu} \,, \qquad J^\mu{}_{ab} =  \frac{\kappa}{e}\frac{\delta S_{{\tiny\mbox{mat}}}}{\delta \omega^{ab}_\mu} \,,\\[.1truecm]
T_\mu &=& \frac{\kappa}{e}\frac{\delta S_{{\tiny\mbox{mat}}}}{\delta \tau^\mu} \,,\qquad  T_\mu{}^a =\frac{\kappa}{e}\frac{\delta S_{{\tiny\mbox{mat}}}}{\delta e_a^\mu} \,.
\end{eqnarray}

\subsection{Matter coupled Carroll gravity}\label{sec: Car matter}

For any $D>2$ the set of equations of motion obtained by varying the action \eqref{total} with respect to $\tau^\mu$, $e^\mu_a$, $\omega_\mu^a$ and $\omega_\mu^{ab}$ 
can be written as follows:
\begin{eqnarray}
R_{ab}(H) &=& 2J^0{}_{ab}\,, \label{Careom1 mat}\\
R_{0b}(H) &=& J^0{}_{b} - \frac{1}{D-2}\left( J^0{}_{b}  + 2J^a{}_{ab}\right)\,,\\
R_{ab}{}^c(P) &=&  \frac{2}{D-2}\left(J^0{}_{[a}\delta^c_{b]}  +2J^d{}_{d[a}\delta^c_{b]}\right)  + 2J^c{}_{ab}\,,\\
R_{a0}{}^c(P) &=&  \frac{1}{D-2}\delta^c_a J^b{}_b - J^c{}_a \,,\label{Careom4 mat}\\
T_0 &=& - \frac{1}{2}R_{ab}{}^{ab}(J)\,,\hskip 1.5truecm
T_a = R_{ab}{}^b(G) \,,\label{Careom5 mat}\\
T_0{}^a &=& R_{0b}{}^{ab}(J) \,,\\
T_a{}^b &=& R_{ac}{}^{bc}(J) -R_{a0}{}^b(G) + \delta^b_a R_{c0}{}^c(G) - \frac{1}{2} \delta^b_a  R_{cd}{}^{cd}(J) \,.\label{Careom7 mat}
\end{eqnarray}
The equations (\ref{Careom1 mat}) - (\ref{Careom4 mat}) can be used to solve for the spin connections
\begin{eqnarray}
\omega_{\mu a} &=& \tau_\mu \tau^\nu e^\rho_a \partial_{[\nu}\tau_{\rho]} + e^\nu_a\partial_{[\mu}\tau_{\nu]} + e^b_\mu S_{ab} - \left(\frac{(D-3)J^0{}_a + 2 J^b{}_{ab}}{D-2}\right)\tau_\mu  + J^0{}_{a b}e^{b}_{\mu}\,, \label{eq: solomegaa mat}\\[.1truecm]
\omega_\mu{}^{ab} &=& - 2 e^{\rho [a}\partial_{[\mu} e_{\rho]}^{b]} + e_{\mu c} e^{\rho a} e^{\nu b}\partial_{[\rho} e_{\nu]}^c  - J^{[ab]}\tau_{\mu} - \frac{2}{D-2}\left(J^0{}_d + 2J^c{}_{cd}\right)\delta^{d[a}e^{b]}_{\mu}  \nonumber \\ [.1truecm] && - 2J^{[a}{}_{cd}\delta^{b]d}e^{c}_{\mu}
- \delta^{ac}\delta^{bd}J^f{}_{cd}e_{f\mu} \,.\label{eq: solomegaab mat}
\end{eqnarray}
The same equations can also be used to derive the  constraint
\begin{equation}
K_{ab} =  J_{(ab)} - \frac{1}{D-2}\delta_{ab} J^c{}_{c}  \,, \label{eq: constraint viel mat}
\end{equation}
on the extrinsic curvature.

Like in the case of General Relativity discussed in section 2 the equations of motion \eqref{Careom1 mat} - \eqref{Careom7 mat} give relations between the curvatures and the currents. The Bianchi identities \eqref{eq: CarBianchi1} - \eqref{eq: CarBianchi4} then lead to the following on-shell relations among the currents
\begin{eqnarray}
T_{0a} &=& - \mathcal{D}_b J^b{}_a  - \mathcal{D}_0 J^0{}_a +\frac{1}{D-2}\left(  2J^b{}_a J^c{}_{bc} + J^0{}_a J^c{}_c  - J^0{}_b J^b{}_a  \right) \,,\\[.1truecm]
T_{[ab]}  &=& - \mathcal{D}_0 J^0{}_{ab}  - \mathcal{D}_c J^c{}_{ab} \nonumber\\[.1truecm] &&  - \frac{1}{D-2}\left(J^0{}_c J^c{}_{ab} - 2J^c{}_{ab}J^f{}_{cf}  - J^c{}_c J^0{}_{ab} \right) \,,\\[.1truecm]
\mathcal{D}_0 T_0 + \mathcal{D}_a T_0{}^a  &=& J^0{}_a T_0{}^a + J^a{}_b R_{ac}{}^{bc}(J) - J^a{}_{b c} R_{0 a}{}^{b c}(J) \nonumber\\[.1truecm] &&+ \frac{2}{D-2}\left( J^a{}_a T_0-  J^0{}_a T_0{}^a  - 2J^a{}_{ab}T_0{}^b \right) \,,\\[.1truecm]
\mathcal{D}_0 T_c +\mathcal{D}_a T_c{}^a &=& - J^\mu{}_c T_\mu  + 2J^\mu{}_{cb} T_\mu{}^b  + J^\mu{}_a R_{\mu c}{}^a(G) + J^\mu{}_{a b} R_{\mu c}{}^{ab}(J) \nonumber\\[.1truecm] &&
\hskip -1truecm + \frac{2}{D-2}\left( \left(J^0{}_c+2J^a{}_{ac}\right) \frac{( T_a{}^a  + T_0 )}{2} - (J^0{}_b  + 2 J^a{}_{ab})T_c{}^b + J^a{}_a T_c\right) \,.
\end{eqnarray}

\subsection{Matter coupled Galilei gravity}

The equations of motion that follow from the Galilei action with matter \eqref{total} for any $D>2$ are given by
\begin{eqnarray}
R_{ab}(H) &=&  2J^0{}_{ab}\,, \label{eq: galeom mat0}\\[.1truecm]
R_{0a}(H) &=&  \frac{1}{D-2}\left( 2J^b{}_{ab} + (D-3) R_{ab}{}^{b}(P) \right)\,, \label{eq: galeom mat1} \\[.1truecm]
R_{ab}{}^{c}(P) &=& 2 J^{c}{}_{ab}+\frac{2}{D-2} \delta_{[a}^c\left(  2J^d{}_{b]d} -R_{b]d}{}^{d}(P) \right)\,,\label{eq: galeom mat2} \\[.1truecm]
T_0 &=& - \frac{1}{2} R_{ab}{}^{ab}(J) \,,\hskip 1.5truecm
T_a = 0 \,, \\[.1truecm]
T_0{}^a &=&  R_{0b}{}^{ab}(J)\,,\\[.1truecm]
T_{ab} &=&  R_{acb}{}^{c}(J) - \frac{1}{2}\delta_{ab}R_{bc}{}^{bc}(J) \,. \label{eq: galeom mat5}
\end{eqnarray}
The fact that $T_a=0$ is a direct consequence of the Galilei boost invariance of the action. Furthermore, the local scale invariance
given by eqs.~\eqref{eq: rescaltau} and \eqref{eq: rescale} implies
\begin{equation}
T_{aa} = (D-3)T_0  \,.
\end{equation}

For $D>3$ the equations of motion \eqref{eq: galeom mat1} and \eqref{eq: galeom mat2} can be used to solve for the spatial rotation spin connection $\omega_\mu{}^{ab}$ as follows
\begin{eqnarray}
\omega_\mu^{ab} &=& \tau_{\mu}A^{a b} + Z^{abc}e_{c\mu}\,, \\
Z_{abc} &=& \left( \delta_{cd} e_{[a}^{\mu} e_{b]}^{\nu} +\delta_{bd} e_{[a}^{\mu} e_{c]}^{\nu} - \delta_{ad}  e_{[b}^{\mu} e_{c]}^{\nu} \right) \partial_{\mu}{e^{d}_{\nu}}
- J^{d}{}_{ab} \delta_{cd} + J^{d}{}_{bc} \delta_{ad} - J^{d}{}_{ac} \delta_{bd} \\ &&
+ \frac{4}{D-3}\left( e_{[a}^{\mu} \delta_{b]c} \tau^{\nu}\partial_{[\mu}\tau_{\nu]}  - J^d{}_{d[a} \delta_{b]c}  \right) \,,
\end{eqnarray}
except for an anti-symmetric tensor component  $A^{a b}=-A^{b a}$ of $\omega_\mu{}^{ab}$.

The case $D=3$ is special. In this case we may write $J^\mu{}_{ab}=\epsilon_{ab}J^\mu$ and the equations \eqref{eq: galeom mat0} and \eqref{eq: galeom mat1} imply the constraint
\begin{equation}
\partial_{[\mu}\tau_{\nu]} = \epsilon_{ab}\left( J^0e^a_\mu e^b_\nu + 2\tau_{[\mu} e^a_{\nu]}J^b\right) \,.
\end{equation}
The current $J^\mu$ automatically drops out from equation \eqref{eq: galeom mat2} which is solved by a fully undetermined spin connection $\omega^{ab}_\mu = \epsilon^{ab} \omega_\mu$.



Finally, like in General Relativity and Carroll gravity, the equations of motion \eqref{eq: galeom mat0} - \eqref{eq: galeom mat5}
give relations between the curvatures and the currents. Using the Bianchi identities \eqref{eq: GalBianchi1} - \eqref{eq: GalBianchi4}
the equations of motion imply the following additional on-shell relations between the currents:
\begin{eqnarray}
T_{[ab]}  &=& - \mathcal{D}_0 J^0{}_{ab} - \mathcal{D}_c J^c{}_{ab} - J^0{}_{a b} R_{0c}{}^c(P) \nonumber \\ && +\frac{1}{D-2}J^c{}_{a b}\left( 2J^d{}_{cd} - R_{cd}{}^{d}(P) \right) \,, \qquad\\[.1truecm]
\mathcal{D}_0 T_0 +  \mathcal{D}_a T_0{}^a &=& - J^a{}_{bc}R_{0 a}{}^{b c}(J) + T_0{}^a R_{ab}{}^b(P)  + T_a{}^b R_{0b}{}^{a}(P)  - T_0 R_{0a}{}^a(P) \nonumber\\&& - \frac{2}{D-2}T_0{}^a  \left(  2J^{b}{}_{ba} + R_{ac}{}^c(P)\right)\,,\\[.1truecm]
\mathcal{D}_b T_{a}{}^{b} &=& J^{\mu}{}_{bc}R_{\mu a}{}^{bc}(J) + 2J^\mu{}_{ab}T_\mu{}^b - 2J^b{}_{ab} T_0 \nonumber \\ && - \frac{2}{D-2}T_a{}^b\left( 2J^c{}_{cb}  + R_{bc}{}^c(P) \right)\,.
\end{eqnarray}

\subsection{Examples}

In the previous section, we have left the matter action unspecified. In this section, we will consider specific examples of matter actions coupled to arbitrary Carrollian and Galilean backgrounds.
In particular, we will consider actions for a real scalar field, a Dirac field and electromagnetism. The starting point in all cases will be the corresponding matter action coupled to a fixed relativistic background. After that, we will study the corresponding Carrollian and Galilean limits.

\subsubsection{Spin 0}

We first consider the action for a real Klein-Gordon
field $\Phi$, with mass $M$, minimally coupled to an arbitrary relativistic background
\be \label{eq:KGaction}
S_{\tiny{\mbox{KG}}}= -\frac 12 \int d^Dx \sqrt{-g}\Big( g^{\mu\nu}\partial_\mu\Phi(x) \partial_\nu
\Phi(x)+M^2 \Phi(x)^2\Big)\,.
\ee
Focusing first on the Carrollian limit, we find that upon applying the rescalings (\ref{eq: rescal1}), (\ref{eq: rescal2}), along with $\Phi=\frac{1}{\sqrt{\omega}} \phi$ and $M = \omega m$, the $\omega \rightarrow \infty$ limit of (\ref{eq:KGaction}) leads to the following Carroll action
\be
S_{\tiny{\mbox{KG}}}^{\tiny{\mbox{Car}}}=\frac 12 \int d^Dx e\Big(\tau^\mu\tau^\nu \partial_\mu\phi(x) \partial_\nu
\phi(x)-m^2 \phi(x)^2\Big) \,.\label{eq: carspin0}
\ee
The equation of motion for $\phi$ is then given by
\begin{equation} \label{eq:carrollspin0eom}
\left( \mathcal{D}^2_0 +  m^2\right)\phi  = 0\,,
\end{equation}
where $\mathcal{D}^2_0=\tau^\mu\partial_\mu (\tau^\nu\partial_\nu)$ is the second order Carroll-covariant time derivative. This equation of motion has appeared in a first order form in \cite{Bergshoeff:2014jla}.

Another way of arguing that equation (\ref{eq:carrollspin0eom}) is the correct equation of motion for a scalar field in an arbitrary Carroll background, is by considering the Carroll limit of a relativistic particle in a relativistic curved background specified by the metric $g_{\mu\nu}$. The canonical action of such a particle is given by

\be\label{relactioncurved}
S=\int d\tau\,\Big[p_\mu\dot{ x^\mu}-\frac \lambda 2 E \Big( g^{\mu\nu}p_\mu p_\nu+M^2\Big)\Big]\,,
\ee
where $E = \det(E_\mu{}^A)$.
The Carrollian limit is obtained by applying the rescalings
(\ref{eq: rescal1}), (\ref{eq: rescal2}), along with $M=\omega m$ and by taking the limit $\omega \rightarrow \infty$. The dominant term is given by
\be\label{actioncurved}
S=\int d\tau\,\Big[p_\mu\dot{ x^\mu}-\frac \lambda 2 e \Big(-\tau^\mu(t,\vec x)\tau^\nu(t,\vec x) p_\mu p_\nu+m^2\Big)\Big]\,,
\ee
where a factor of $\omega$ has been absorbed in $\lambda$.
The equations of motion obtained by varying the coordinates and momenta are given by
\be\label{move2}
\dot x^\mu= - e\lambda\tau^\mu\tau^\nu p_\nu,\qquad \dot p_\mu= e \lambda(\partial_\mu\t^\rho)\tau^\sigma p_{\sigma} p_{\rho}\,.
\ee
By varying with respect to the Lagrange multiplier $\lambda$, one obtains the mass-shell constraint for a Carroll particle
\be
-\tau^\mu(t,\vec x)\tau^\nu(t,\vec x) p_\mu p_\nu+m^2 = 0 \,.
\ee
Upon quantization, i.e.~replacing $\tau^\mu p_\mu \rightarrow - i \mathcal{D}_0$, this mass-shell constraint indeed leads to the equation of motion (\ref{eq:carrollspin0eom}) of a spin 0 field.

In the Galilean case we perform the same rescaling on the scalar field, $\Phi = \omega^{-\frac{1}{2}}\phi$, but we keep the mass $M$ as it is. We thus obtain
\begin{equation}
S_{\tiny{\mbox{KG}}}^{\tiny{\mbox{Gal}}} = - \frac{1}{2} \int e \left( \delta^{ab}e_a^\mu e_b^\nu\partial_\mu \phi\partial_\nu \phi + M^2\phi^2 \right) \,. \label{eq: galspin0}
\end{equation}
The equation of motion for $\phi$ is given by
\begin{equation}
\left(\delta^{ab}\mathcal{D}_a\mathcal{D}_b + \frac{1}{D-2}R_{ab}{}^{b}(P)\mathcal{D}_a -  M^2\right)\phi   = 0\,,
\end{equation}
where $\mathcal{D}_a\mathcal{D}_b\phi=e_a^\mu(\partial_\mu \mathcal{D}_b \phi - \omega^{bc}_\mu\mathcal{D}_c\phi)$ is the second order Galilean-covariant spatial derivative. Written as such this result is valid for any $D\neq2$. For $D>3$ we have the additional relation $R_{ab}{}^{b}(P) \propto 2\partial_{[\mu}\tau_{\nu]}\tau^\mu e^\nu_a = R_{0a}(H)$.

\subsubsection{Spin $\frac12$}
We now consider the coupling of a Dirac field to a curved background.\footnote{The case of a spinning particle coupled to a curved background could also be studied.}
The action is given by
\be
S_{\mathrm{Dirac}}=  \int d^Dx \sqrt{-g}\, \bar\Psi\gamma^\mu\left( \partial_\mu-\frac 14 \Omega_\mu^{AB}\gamma_{AB}\right)\Psi \,,
\ee
where $\gamma^\mu = E^\mu{}_A \gamma^A$.

Using the rescalings (\ref{eq: rescal1}) and (\ref{eq: rescal2}) and taking the limit $\omega\rightarrow \infty$, one finds the following `Carroll-Dirac' action
\be
S_{\mathrm{Carroll-Dirac}}=  \int d^Dx e \, \bar\Psi\gamma^0\tau^\mu\left( \partial_\mu-\frac 14 \omega_\mu^{ab}\gamma_{ab}\right)\Psi\,.
\ee
As in the scalar field case, this action only contains a time-like derivative. Furthermore, it is interesting to note that it only contains the spin connection $\omega_\mu{}^{ab}$ that does not contain any undetermined components.

The Galilean limit is obtained by applying the rescalings (\ref{eq: Galrescal1}), (\ref{eq: Galrescal2}) and $\Psi \rightarrow \frac{1}{\sqrt{\omega}} \Psi$ and taking the limit $\omega\rightarrow \infty$. This leads to the `Galilei-Dirac' action
\be
S_{\mathrm{Galilei-Dirac}}=  \int d^Dx e \, \bar\Psi\gamma^a e^\mu{}_a \left( \partial_\mu-\frac 14 \omega_\mu^{bc}\gamma_{bc}\right)\Psi\,.
\ee
Like for the scalar field, this action only contains a spatial derivative. It also contains the spin connection $\omega_\mu{}^{ab}$. It does however not contain the undetermined components of the latter, as these components lie along the $\tau_\mu$ direction and are projected out of the above action since $\omega_\mu{}^{bc}$ appears multiplied with $e^\mu{}_a$.

Unlike the Carroll case, in the Galilean case one could consider a different limit, with different components of the fermion scaling differently,  that does lead to the appearance of a (undetermined) boost connection field in the action and fermions that transform under Galilean boosts. This other limit is basically the massless limit of the Newton-Cartan limit considered in \cite{Bergshoeff:2015sic}, see eq.~(2.6) of that paper.

\subsubsection{Spin 1: Electromagnetism}

Starting from the action for Maxwell electromagnetism coupled to an arbitrary relativistic background
\begin{equation}
  \label{eq:maxwellaction}
  S_{\mathrm{Maxwell}} = - \frac14 \int d^Dx \sqrt{-g}\, g^{\mu\rho} g^{\nu \sigma} F_{\mu\nu} F_{\rho \sigma} \,,
\end{equation}
with $F_{\mu\nu} = 2 \partial_{[\mu} A_{\nu]}$, the Carrollian limit is obtained by applying the rescalings (\ref{eq: rescal1}), (\ref{eq: rescal2}) of the background fields, along with a rescaling $A_\mu \rightarrow \frac{1}{\sqrt{\omega}} A_\mu$ and by taking the limit $\omega \rightarrow \infty$. In this way, one obtains the following `Carroll-Maxwell' action
\begin{equation}
  \label{eq:carrollmaxwell}
  S_{\mathrm{Carroll-Maxwell}} = \frac12 \int d^Dx\, e\, \left(\tau^\mu F_{\mu\nu}\right) \left(\tau^\rho F_{\rho\sigma}\right) h^{\nu \sigma} \,.
\end{equation}
Similarly, the Galilean limit is obtained by taking the limit $\omega \rightarrow \infty$, after applying the rescalings (\ref{eq: Galrescal1}), (\ref{eq: Galrescal2}) and $A_\mu \rightarrow \frac{1}{\sqrt{\omega}} A_\mu$. This leads to the `Galilei-Maxwell' action
\begin{equation}
  \label{eq:galileimaxwell}
  S_{\mathrm{Galilei-Maxwell}} = -\frac14 \int d^Dx \, e\, h^{\mu\rho} h^{\nu\sigma} F_{\mu\nu} F_{\rho\sigma} \,.
\end{equation}
One thus sees that the Carroll-Maxwell Lagrangian is the generalization of $\vec{E}\cdot \vec{E}$ to arbitrary Carroll backgrounds, where $\vec{E}$ is the electric field.\footnote{When restricted to flat space-time, the Carroll-Maxwell action corresponds to the action of `Carrollian electromagnetism of the electric type', considered in \cite{Duval:2014uoa} and more recently in \cite{Bagchi:2016bcd} in the context of flat space holography. In \cite{Duval:2014uoa}, `Carrollian electromagnetism of the magnetic type' is also considered, whose Lagrangian is given by $\vec{B}\cdot\vec{B}$. This theory can, however, be obtained from Carrollian electromagnetism of the electric type, by interchanging $\vec{E} \rightarrow \vec{B}$ and $\vec{B} \rightarrow -\vec{E}$.} Similarly, the Galilei-Maxwell Lagrangian is a suitable generalization of $\vec{B}\cdot \vec{B}$, with $\vec{B}$ the magnetic field, to arbitrary Galilean backgrounds. While it may seem puzzling at first that only the electric field appears in the Carroll-Maxwell Lagrangian, this is consistent with the fact that the dynamics of Carroll particles and fields is trivial, in the sense that their equations of motion only involve time derivatives. As a consequence, minimal coupling to a vector potential will only involve the electric potential. Physically, since Carroll particles do not move, they will not induce a magnetic field nor will they be subjected to a Lorentz magnetic force. It therefore makes sense that the Carroll-Maxwell Lagrangian only involves the electric field, as that is the only field that will be relevant in coupling to Carroll particles and fields.

Similarly, actions for Galilei fields only involve spatial derivatives and minimal coupling to a vector potential will likewise only involve the spatial parts of the vector potential. The Galilei-Maxwell action then only contains the magnetic field, as that is the only contribution relevant for couplings to Galilei fields.

Note that the Galilei-Maxwell action above does not correspond to the action of what is known in the literature as Galilean electrodynamics \cite{leBellac} (for a review, see \cite{Rousseaux}; see \cite{Bagchi:2014ysa,Bagchi:2015qcw} for a discussion in the context of flat space holography), coupled to an arbitrary non-relativistic background. The latter contains contributions from both the electric and magnetic fields. While this action can not be obtained via the simple limit considered in this paper, it can be obtained by taking different limit procedures. In particular, it arises as a non-relativistic limit of an action that is a sum of the Maxwell action and the action for a real massless scalar field in an arbitrary relativistic background \cite{Bergshoeff:2015sic}. The action for Galilean electrodynamics in flat space-time has also been obtained via null reduction in \cite{Festuccia:2016caf}. As Galilean electrodynamics involves both electric and magnetic fields, it is the appropriate theory to consider when dealing with non-relativistic charged particles and fields, whose equations of motion involve both spatial and time derivatives. Examples of such fields have been studied in \cite{Bergshoeff:2015sic}. These examples involve massive fields and exhibit mass conservation. The appropriate non-relativistic background to couple such fields to is then a Newton-Cartan background, which we mentioned in the introduction. This Newton-Cartan background is an extension of a Galilean background, that apart from $\tau_\mu$ and $e_\mu{}^a$ also involves an extra one-form $m_\mu$, that plays the role of gauge field associated to the charge that expresses mass conservation.

\section{Conclusions}

In this paper we showed that there exist two consistent limits of the Einstein-Hilbert action describing General Relativity that lead to finite actions, upon making a redefinition of Newton's constant. This is in contrast to the Newton-Cartan limit, leading to Newton-Cartan gravity,  that we defined in \cite{Bergshoeff:2015uaa} and that can be taken at the level of the equations of motion only. The first, ultra-relativistic, limit leads to a so-called Carroll gravity action while the second limit is non-relativistic and leads to a so-called Galilei gravity action. We presented the actions both in first-order and second-order form. A noteworthy feature is that, unlike General Relativity, not all components of the spin connection fields can be solved for. We showed that the independent components occur as Lagrange multipliers in the action thereby imposing constraints on the geometry. The case of Carroll gravity is interesting in view of possible applications to flat space holography where the Carroll symmetries play an important role \cite{Banks:2016xsj}.

Here, we have considered Carrollian and Galilean limits of General Relativity at the level of the action. One could also consider these limits at the level of the equations of motion. However, this is not an unambiguous procedure. The relativistic equations of motion that one starts from can be written in different equivalent ways, that can however lead to different limits when $\omega\rightarrow\infty$. For instance, the limit taken directly in \eqref{eq: varS Omega} (with $A = a$ and $B=0$) is divergent in the Carroll case but not in Galilean one. On the other hand, the limit in the same equation of motion rewritten simply as $R_{\mu\nu}{}^A(P)=0$ is divergent in the Galilean case and not in the Carroll one. It would be interesting to further investigate the possible limits of the equations of motion.

Given pure General Relativity, without additional fields, the Carroll and Galilei limits are the only consistent ones that can be taken at the level of the Einstein-Hilbert action.\footnote{Using an ADM formulation one can additionally define the  strong coupling limit of \cite{Henneaux:1979vn,Henneaux:1981su}.}
Using an expansion of the fields in terms of the contraction parameter $\omega$ this limit picks out the leading term in an $\omega$-expansion of the action. Introducing an additional vector field, a (non-relativistic) Newton-Cartan limit at the level of the equations of motion can be defined leading to the equations of motion of Newton-Cartan gravity. From the $\omega$-expansion point of view, the vector field helps in cancelling the leading (divergent) term in an $\omega$-expansion of the equations of motion with the effect that this new non-relativistic limit picks out the (finite) subleading term in an $\omega$-expansion. It would be interesting to see whether, using the same vector field, also an ultra-relativistic limit can be defined that picks out the subleading term in the $\omega$-expansion and whether the resulting `Carroll gravity' theory is related to the one presented in \cite{Hartong:2015xda}.

After constructing the gravity actions, we also considered matter couplings and compared the results with the case of Newton-Cartan gravity. A characteristic feature of these matter couplings is that only time derivatives (Carroll limit) or spatial derivatives (Galilei limit) survive whereas in a Newton-Cartan limit both types of derivatives survive like in the case of the Schr\"odinger action. In the case of spin 0 Carroll matter, we showed that the results obtained  are consistent with the point of view of a Carroll particle.

Besides taking the Carroll or Galilei limit of General Relativity, one could also consider taking these limits at the level of the effective actions that describe extended objects beyond particles. For instance, Carroll strings have been considered in \cite{Cardona:2016ytk}. Recently, a Galilean limit of a relativistic Green-Schwarz superstring action has been considered and the resulting non-relativistic so-called Galilean superstring, exhibiting kappa-symmetry,  has been given \cite{Gomis:2016zur}. One could also consider `stringy' versions of the limits considered in this paper where, besides the time direction, one or more of the spatial directions, those in the direction of the world-volume of the extended object, play a special role.

It would be interesting to apply the Hamiltonian canonical quantisation procedure to Carroll and Galilei gravity and verify how many physical degrees of freedom exist in these models. This would enable one to find out whether the Lagrange multiplier fields do represent any kind of non-relativistic degree of freedom.

In a previous paper \cite{Bergshoeff:2016soe} we already discussed the extension of this work to include higher spins, i.e.~fields describing particles with spin larger than 2. It would be interesting to see whether the geometries discussed in \cite{Bergshoeff:2016soe} have  applications to the non-relativistic higher-spins that have recently been discussed in the context of the fractional quantum Hall liquid \cite{Golkar:2016thq} in the same way as Newton-Cartan geometry has  found applications in Condensed Matter Theory, see, e.g., \cite{Hoyos:2011ez}.

\section*{Acknowledgements}
J.G.~and T.t.V.~acknowledge the hospitality at the Van Swinderen Institute for Gravity and Particle Physics
of the University of Groningen where most of this work was done. E.B., J.G.~and J.R. thank the GGI in Firenze for the
stimulating atmosphere during the workshop {\it Supergravity: what next?} when part of this work was done.
J.G has been supported  in part by FPA2013-46570-C2-1-P, 2014-SGR-104 (Generalitat de Catalunya) and Consolider CPAN and by
the Spanish goverment (MINECO/FEDER) under project MDM-2014-0369 of ICCUB (Unidad de Excelencia Mar\'\i a de Maeztu).

\appendix

\end{document}